# Néel-type skyrmions and their current-induced motion in van der Waals ferromagnet-based heterostructures

**Authors**: Tae-Eon Park,[1] Licong Peng,[2] Jinghua Liang,[3] Ali Hallal,[4] Fehmi Sami Yasin,[2] Xichao Zhang,[5] Kyung Mee Song,[1] Sung Jong Kim,[1,6] Kwangsu Kim,[1,7] Markus Weigand,[8] Gisela Schütz,[9] Simone Finizio,[10] Jörg Raabe,[10] Karin Garcia,[11] Jing Xia,[5] Yan Zhou,[5] Motohiko Ezawa,[12] Xiaoxi Liu,[13] Joonyeon Chang,[1,14] Hyun Cheol Koo,[1,6] Young Duck Kim,[15] Mairbek Chshiev,[4] Albert Fert,[11,16] Hongxin Yang,[3*] Xiuzhen Yu,[2*] Seonghoon Woo[1,17*]

**Affiliation.**

[1]Center for Spintronics, Korea Institute of Science and Technology, Seoul 02792, Korea

[2]RIKEN Center for Emergent Matter Science, Wako, 351-0198, Japan

[3]Ningbo Institute of Materials Technology and Engineering, Chinese Academy of Sciences, Ningbo 315201, China

[4]SPINTEC, CEA/CNRS/UFG-Grenoble 1/Grenoble-INP, INAC, 38054 Grenoble, France

[5]School of Science and Engineering, The Chinese University of Hong Kong, Shenzhen, Guangdong 518172, China

[6]KU-KIST Graduate School of Converging Science and Technology, Korea University, Seoul 02841, Korea

[7]Department of Physics, University of Ulsan, Ulsan 44610, Korea

[8]Helmholtz-Center Berlin, Albert-Einstein-Straβe 15, 12489 Berlin, Germany

[9]Max-Planck-Institut für Intelligente Systeme, 70569 Stuttgart, Germany

[10]Swiss Light Source, Paul Scherrer Institut, 5232 Villigen, Switzerland

[11]DIPC and University of the Basque Country, 2018, San Sebastian, Spain

[12]Department of Applied Physics, University of Tokyo, Hongo 7-3-1, Tokyo 113-8656, Japan

[13]Department of Electrical and Computer Engineering, Shinshu University, Wakasato 4-17-1, Nagano 380-8553, Japan

[14]Department of Materials Science & Engineering, Yonsei University, Seoul 03722, Korea

[15]Department of Physics, Kyung Hee University, Seoul 02447, Korea

[16]Unité Mixte de Physique, CNRS, Thales, Univ. Paris-Sud, Université Paris-Saclay, Palaiseau 91767, France

[17]IBM T.J. Watson Research Center, 1101 Kitchawan Rd, Yorktown Heights, New York 10598, USA

* Correspondence to: hongxin.yang@nimte.ac.cn, yu_x@riken.jp, shwoo@ibm.com

**Since the discovery of ferromagnetic two-dimensional (2D) van der Waals (vdW) crystals, significant interest on such 2D magnets has emerged, inspired by their appealing properties and integration with other 2D family for unique heterostructures. In known 2D magnets, spin-orbit coupling (SOC) stabilizes perpendicular magnetic anisotropy (PMA). Such a strong SOC could also lift the chiral degeneracy, leading to the formation of topological magnetic textures such as skyrmions through the Dzyaloshinskii-Moriya interaction (DMI). Here, we report the experimental observation of Néel-type chiral magnetic skyrmions and their lattice (SkX) formation in a vdW ferromagnet $Fe_3GeTe_2$ (FGT). We demonstrate the ability to drive individual skyrmion by short current pulses along a vdW heterostructure, FGT/*h*-BN, as highly required for any skyrmion-based spintronic device. Using first principle calculations supported by experiments, we unveil the origin of DMI being the interfaces with oxides, which then allows us to engineer vdW heterostructures for desired chiral states. Our finding opens the door to topological spin textures in the 2D vdW magnet and their potential device application.**

Two dimensional (2D) van der Waals (vdW) crystals have been significantly highlighted as a unique material platform, mainly due to their fascinating physical properties, low-cost fabrication and high integrability to produce appealing artificial heterostructures[1,2]. Recent addition of magnetic vdW crystals, where intrinsic long-range magnetic orders were observed in $Cr_2Ge_2Te_6$ and $CrI_3$, offered a new building block to this platform, opening a whole new door to vdW magnet-based spintronics[3–10]. Therefore, significant following interests have emerged and rapidly demonstrated few key elements for applications, including the magnetoresistance (MR) effects[5,6] and gate-tunable room-temperature magnetism[7].

Whereas the long-range magnetic order is often suppressed in vdW crystals due to thermal fluctuations given by Mermin-Wagner theorem[11], strong spin-orbit coupling (SOC) in vdW magnets plays an essential role in stabilizing the perpendicular magnetic anisotropy (PMA) and thus overcomes the thermal fluctuations down to a monolayer limit[4,7]. In a material with such large SOC and broken inversion symmetry, the anti-symmetric exchange interaction, so called Dzyaloshinskii-Moriya interaction (DMI)[12,13], can emerge and be strong enough to stabilize topological magnetic configurations including skyrmions[14,15]. Recent theoretical works have also discussed the emergence of strong DMI in vdW magnets with various possible origins, e.g. crystal symmetry or material boundary, as well as resulting skyrmion stabilization[16–18]. Once established, by taking advantages of other 2D crystals that are stackable and offer unique electrical properties[1,2], this material platform could provide a novel route towards skyrmion-based devices that have been challenging with conventional metallic ferromagnets[19,20]. However, experimental demonstration of such chiral magnetic skyrmions and their current-driven dynamics in vdW magnets and heterostructures has remained elusive so far.

Here, we experimentally present the observation of Néel-type skyrmions and their ordered crystal lattice structures in vdW ferromagnetic $Fe_3GeTe_2$-based heterostructures (FGT hereafter). Moreover, we demonstrate the ability to drive skyrmions using nanoseconds current-pulses in 2D heterostructure based on FGT, i.e. FGT/*h*-BN. Among various types of vdW magnets, FGT exhibits relatively high ferromagnetic transition temperature ($T_C$), large PMA, and metallic nature that enables efficient charge/spin transport suitable for spintronic applications[7,21]. In this study, we utilize first high spatial resolution magnetic imaging techniques, scanning transmission X-ray microscopy (STXM), Lorentz transmission electron microscopy (LTEM) and differential phase contrast microscopy (DPCM) to directly observe magnetic structures in the FGT-based heterostructures. We first show the dynamic generation and stabilization of multiple skyrmions and their crystal (SkX, also referred to as skyrmion lattice) state in FGT, where strong pulse-induced thermal fluctuations transform magnetic domains into SkX. We then examine the stability of SkX against thermal fluctuation and magnetic fields, which eventually constitutes experimental phase diagram of SkX state. We also present the static generation of magnetic skyrmions and SkX using a tilted magnetic field, where we simultaneously unveil the Néel-type chiral nature of skyrmions stabilized in the SkX state by taking advantage of in-plane magnetization sensitivity in LTEM measurements. Moreover, we demonstrate the current-driven motion of skyrmions in FGT-based vdW heterostructure (FGT/*h*-BN), where we drive isolated individual skyrmions by short current pulses along a vdW heterostructure racetrack at speeds approaching a meter per second. Finally, using first principle calculations corroborated by additional DPCM measurements, we prove the presence of significant interfacial DMI at FGT interfaces with oxidized layers, which then allows us to 'engineer' the chiral states in vdW ferromagnetic heterostructure between Bloch and Néel-types, by using selective fabrication processes.

Figure 1a schematically shows the crystal structures of mono-layered FGT viewed from *xy* and *yz* planes and bi-layered FGT exhibiting vdW bonding between monolayers. Each FGT monolayer consists of a Fe$_3$Ge covalently bonded slab and two Te layers placed above and underneath the Fe$_3$Ge, and each layer is separated by a 2.95 Å vdW gap in multi-layered stack[22]. Within a Fe$_3$Ge slab, two inequivalent Fe sites exist, Fe$^{II}$ (the valence states of Fe$^{2+}$) and Fe$^{III}$ (the valence states of Fe$^{3+}$). Overall, the reduced bulk crystal symmetry in FGT is known to provide a magnetocrystalline anisotropy induced by strong SOC[23]. For both electrical and transmission microscopy measurements on the same sample, we fabricated Hall-bar type FGT devices on a 100-nm-thick Si$_3$N$_4$ membrane using dry mechanical exfoliation technique together with e-beam lithography and lift-off (see Methods and Supplementary Fig. S1 for details). Figure 1b shows the cross-sectional view of high-resolution transmission electron microscopy (HRTEM) images of the device, where layered high-crystalline quality FGT is observed (Fig. 1b, inset). Note that FGT layer is sandwiched by two oxidized FGT (O-FGT) due to the sample fabrication under ambient condition, and 5 nm-thick Pt was deposited *ex-situ* as a capping material to prevent further oxidation [see Methods and Supplementary Fig. S2 for details]. The magnetic hysteresis behaviors of the FGT device were measured using the Hall resistance ($R_{xy}$) measurement, where external magnetic field was applied to out-of-plane direction at controlled temperatures ranging 100 - 220 K (Fig. 1c). While the $R_{xy}$ consists of a normal Hall resistance ($R_N$) and an anomalous Hall resistance ($R_{AH}$), FGT films exhibit a large value of $R_{AH}$ in $R_{xy}$, which roughly scales with the magnetization ($M_z$)[7]. Therefore, the square hysteresis loops at 100 K in Fig. 1c corresponds to an out-of-plane magnetic anisotropy, which persists up to 200 K ($T_C \sim 200$ K). It is noteworthy that $R_{xy}$ measurements yields two distinct slopes (sharp and slanted slopes) in temperature range, 100 K $\leq T \leq$ 180 K, and the slanted area becomes more prominent as temperature increases. This area indicates the presence of multi-domain state, which is used to drive the magnetization into multi-domain state at low temperatures and near zero magnetic fields, as discussed in Supplementary Fig. S3. Figure 1d shows the schematic of STXM experimental setup, where the temperature of cooling stage was controlled between 100 K ≤ $T$ ≤ 300 K using liquid nitrogen (LN$_2$) and heat-exchanger. The scanning electron microscopy (SEM) image of measured FGT device with Hall-cross geometry and the electrical circuit diagram is also included in Fig. 1d [see Methods for details]. The magnetization state of FGT device was imaged by probing the intensity of transmitted circularly-polarized X-ray at the Fe-edge ($L_3$ absorption edge), where X-ray magnetic circular dichroism (XMCD) provides contrasts corresponds to the out-of-plane magnetization. Figure 1e shows the magnetic domain configurations in the FGT device as a function of out-of-plane magnetic field, $B_z$, at 120 K, where the dark and bright contrasts in STXM images correspond to downward (-$M_z$) and upward (+$M_z$) out-of-plane magnetization direction of Fe atoms in FGT, respectively. With increasing out-of-plane field $B_z$ > 0, the up domains expand while the down domains shrink into narrow domains, vanishing at the saturation field of $B_z$ = +80 mT.

    Having established that multi-domain states can be readily stabilized and observed in FGT, we then examined the current-induced generation of magnetic skyrmions, as summarized in Fig. 2. In our previous study using conventional chiral ferromagnetic multi-layers, Pt/CoFeB/MgO, we demonstrated that the application of bipolar pulses could transform labyrinth domains with chiral domain walls into multiple skyrmions[24], and the recent study by Lemesh *et al.* [ref.[25]] unveiled the mechanism to be current-induced thermal transformation into skyrmions, because the energy barrier towards the global skyrmionic ground state decreases with increasing temperature. To utilize the same technique on the FGT device, we applied the burst of 100 bipolar pulses, where the pulse frequency of 1 MHz, the peak-to-peak voltage of $V_{pp}$ = 2.96 V and the pulse width of 10

ns were used at $B_z$ = -40 mT and 120 K. As shown in Fig. 2a, it is obvious that the bipolar pulse injection transformed the labyrinth random domain state into multiple circular domain state, where these circular domains turn out to be Néel-type chiral magnetic skyrmions in Fig. 3. We performed the same procedure at slightly lower temperature, 100 K, and the consistent transformation into multiple skyrmions is observed and the generated skyrmions remain stable at zero magnetic field, $B_z$ = 0 mT (highlighted in a blue-boxed area in Fig. 2a). As was observed in ferromagnetic chiral multi-layers, the thermal excitation induced by the bipolar pulses may have opened a path towards global skyrmionics state[24,25]. We examined and observed the consistent domain transformation in another sample capped by graphite (see Supplementary Fig. S4 for details). This demonstration with graphite-capping is significant, as it excludes two possible contributions from Pt: i) the spin-orbit torques (SOTs) by transmitted spin current caused by the spin-Hall effect (SHE) in Pt[26] and ii) the DMI contribution from Pt/O-FGT interface. Additional Hall measurements presented in Supplementary Fig. S5 also confirm the negligible influence from capping materials on the magnetic properties of studied FGT structure. Further analysis reveals that the average size of zero-field skyrmion is ~ 123 nm at 120 K, and the size decreases down to ~ 80 nm with increased density at 160 K [see Supplementary Fig. S6 for details].

At such disordered multi-skyrmion state at 100 K, we applied alternative positive and negative magnetic fields with increasing magnitude up to $B_z$ = ±80 mT with the step of $B_z$ = ±10 mT, as the application of static fields could annihilate pinned weak skyrmions and rearrange them driven by inter-skyrmion repulsive forces, leading to the stabilization of ordered skyrmion state[27,28]. Figure 2b shows the zero-field magnetic configuration after the field sweep, and surprisingly, the initial *disordered* magnetic skyrmions transformed into *ordered* hexagonal SkX. The inset of Fig. 2b presents the enlarged STXM images at a magnetic field, $B_z$ = -80 mT, where the ordered SkX state is more clearly observable (few SkXs are highlighted with blue colors and white lines for guide). The symmetry of SkX also agrees with the symmetry observed in non-centrosymmetric B20-type chiral magnets[14,15]. After stabilizing the SkX state, we then plotted the experimental phase diagram of magnetic configurations in FGT, based on the real-space STXM measurements as summarized in Fig. 2c. We observed three magnetic configuration phases: i) SkX, ii) the co-existence of SkX and multi-domains, and iii) saturated ferromagnetic states, where the representative STXM images of each state are included in the right panel of Fig. 2c. It should be noted that, once generated, SkX in FGT can be stabilized at a wide range of magnetic field (including zero-field) and temperature. Together with the recent discovery of gate-tunable room-temperature magnetization in the same material[7], this observation suggests that it might be possible to harness and manipulate magnetic skyrmions and their lattice at room temperature and zero magnetic fields, which will constitute a major step towards room-temperature skyrmion applications based on vdW magnets.

To unveil the chiral nature of magnetic skyrmions observed by STXM measurements, we then performed the LTEM measurement as summarized in Fig. 3 (see Methods for details). Note here that Fresnel-LTEM is useful to detect the in-plane components of Bloch-type spin configurations at defocused modes, whereas it cannot directly observe Néel-type magnetic configurations with zone-axis beam irradiation, due to the cancellation of magnetic inductions between electrons and symmetric in-plane magnetic moments with opposite directions projected by Néel-type spin textures[29–31]. However, when samples are tilted away from the zone-axis, the projected configurations of up-down magnetic domains should contribute to the LTEM contrasts at defocused modes, therefore, Néel-type magnetic configurations can be observed[28–31]. Figure 3a first shows in- and de-focus LTEM images of the FGT sample tilted about -20° along the *x*-axis at zero field and 160 K, where dark/bright contrasts are only visible in defocused images. Moreover, as shown in the red-boxed areas in the left and right images in Fig. 3a, under- and over-focus LTEM images exhibit the labyrinth domain structures with reserved domain wall contrasts, evidencing the formation of Néel-type chiral multi-domain state in the FGT[30–32]. To generate SkX, we then performed the field cooling (FC) of FGT with an oblique magnetic field of $B$ = -40 mT (the oblique angle is 20° to the zone-axis). Figure 3b shows LTEM images observed at 160 K. Noticeably, the FC generated quasi-static (metastable) Néel-type chiral SkX state in FGT, which magnetic configurations as well as in-plane magnetization profiles agrees well with simulated LTEM results shown in Fig. 3c (see Methods and Fig. S7 for details). Considering that the oblique field applied during the LTEM measurements could have contributed to the in-plane alignment of magnetic moments within FGT domain walls, we have performed LTEM observations of SkX at zero field after the same FC process as described above, which confirmed the robustness of FC-driven Néel-type SkX formation (see Supplementary Fig. S8 for details).

To further highlight the potential of FGT-based 2D vdW heterostructures for skyrmion devices, we next demonstrate the current-driven motion of individual skyrmion in a stacked FGT/$h$-BN vdW bilayer structure, as summarized in Fig. 4. Figure 4a shows a schematic image of the heterostructure track and electric contacts fabricated on $Si_3N_4$ membrane for STXM measurements. Note that a thin $h$-BN vdW flake is used as a capping material on FGT, where the FGT layer is actually sandwiched by two naturally O-FGT as was also shown in Fig. 1b. In this experiment, we first generated initial few-skyrmion-state (as shown in the first image of Fig. 4b) by applying external magnetic field to the multi-skyrmion state acquired by the current-driven skyrmion generation process described in Fig. 2, while a magnetic field of $B_z$ = -50 mT was applied at $T$ = 100 K. In Fig. 4b, each image was obtained after injecting 5 current pulses with $J_a = 1.4 \times 10^{11}$

A/m$^2$ and $t_{\text{pulse}}$ = 50 ns. The current was applied along the +*x* direction, opposite to the electron flow along -*x* direction as schematically indicated in each image.

It is first noteworthy that skyrmions move upon the application of current pulses, and the propagation direction is along the electron-flow direction (against current flow), where this directionality indicates that skyrmion is driven by spin-transfer torques (STTs) arising within the FGT. This is also expected from the HRTEM image shown in Fig. 4a, exhibiting no possible interface of FGT that could provide vertical spin current by e.g. SHE. As shown in Fig. 4c, the average skyrmion velocity was measured to be ~ 1 m/s at a current density $J_e$ = 1.4×10$^{11}$ A/m$^2$, below which no skyrmion motion is observed. The current-driven motion of skyrmion shows the potential of using skyrmions in FGT-based 2D heterostructures for functional device applications, such as the racetrack-type memory[19], where skyrmions act as moveable information carriers. It should also be noted that, although we initially observed three stabilized skyrmions (Sk1-3), only one skyrmion (Sk2) remains stable and propagates along the track during the motion. We speculate such non-ideality may come from the interplay between various magnetic parameters in FGT that sharply changes with temperature, which then alters skyrmion stability as was experimentally navigated in Fig. 2c. However, we believe that further experimental studies of skyrmion motion in FGT-based 2D heterostructures using more efficient torques, e.g. SOTs, arising from some known 2D materials with very large charge-to-spin conversion efficiency, e.g. WTe$_2$ or NbSe$_2$, could substantially improve their electrical controllability and current-velocity relation that could exceed conventional metallic systems[33–36].

With these experimental demonstrations of Néel-type chiral skyrmions, SkX state and their current-driven motion in FGT-based heterostructures, let us now discuss the physical origins of DMI in these structures. We first examined the possible DMI sources from the FGT crystal *symmetry*. As discussed earlier, a monolayer of FGT contains three Fe sublayers (i.e., a 2D Fe$^{II}$ and Ge sublayer between two 2D Fe$^{III}$ sublayer), forming a hexagonal structure that is sandwiched between two Te layers. The whole FGT monolayer structure has the non-centrosymmetric point group of $D_{3h}$[22,37] and thus in principle, shows no bulk DMI. Indeed, although some locally broken inversion symmetry of sublayers in FGT could result in the DMI that stabilizes Néel-type skyrmions, all possible DMI contributions in the whole FGT monolayer structure cancel each other as discussed and summarized in Supplementary Fig. S9 and Supplementary Table S1. For example, similar to the case of 2D hexagonal boron nitride structure with buckling[38], the top Fe$^{III}$ sublayer and neighboring Te layer form a lattice of $C_{3v}$ point group with broken inversion symmetry, the interfacial DMI could be induced at the top Fe$^{III}$ sublayer via the superexchange along the Fe$^{III}$-Te-Fe$^{III}$ path. However, due to the reflection symmetry of the system, the DMI contributions induced at the top and bottom Fe$^{III}$ sublayers are cancelled with each other and the net DMI in the whole FGT structure vanishes.

In order to elucidate the possible origin of DMI at *atomic level*, we performed first principles calculations employing the approach used for multi-layers comprising magnetic and heavy metals[39], oxides[40] and graphene[41] (see Methods for details). We first verified that the DMI for symmetric FGT structure indeed vanishes as discussed above (see Supplementary Fig. S10 for details). For FGT crystal monolayer, the calculated DMI arising at both Fe/Te interfaces is of almost equal magnitude with opposite sign yielding negligible DMI as expected from aforementioned crystal symmetry analysis. This is in agreement with SOC energy difference associated with the total DMI, $\Delta E_{SOC}$, for the same FGT crystal monolayer as shown in Supplementary Fig. S10. We would also like to note that this result of non-existing interfacial DMI in FGT bulk crystal has also been reported recently by Laref et al.[42], opposing to another work by Wang et al.[43] that estimates some finite DMI within FGT.

We next investigated other possible mechanisms of induced DMI by examining global and local atomic distributions along FGT crystal and its interfaces. Figure 5a shows atomic concentrations across the sample acquired using the quantitative high-angle annular dark field detector (HAADF) installed in scanning transmission electron microscopy (STEM) (see Supplementary Fig. S2 for elemental mapping images). One can note first that the concentration of each atom along the thickness of sample within the pure FGT crystal region is symmetric and homogeneous, implying that the DMI owing to asymmetric distribution of elemental content in bulk material, as presented in ref.[44], can be excluded in this case. However, it is noteworthy that there exists significant atomic concentration fluctuation at two FGT/O-FGT interfaces. In particular, the concentration of Te atom at both interfaces rapidly decreases and vanishes upon oxidation, while Fe and Ge concentrations only fluctuate and recover their original values near the largest oxidation areas (oxygen peaks). Figure 5b shows the relative atomic concentration distribution between Te and O atoms, where their sum and difference are plotted. It becomes clearer that, while their total concentration (Te+O) fluctuates within 10 ~ 15%, their concentration difference (Te-O) rapidly decreases from the initial bulk value of Te, ~ 30%, to its negative value, ~ -30%, around oxygen concentration peaks. This distribution variation between Te and O elemental contents strongly implies that Te atoms are likely substituted by O atoms, forming $Fe_3GeTe_{(2-x)}O_x$ over few nanometers of oxidized interfacial regions. Furthermore, one cannot exclude oxygen addition scenario at the interfaces as well. Therefore, we performed systematic calculations of microscopic and micromagnetic DMI parameters ($d$ and $D$), for both O-substitution and O-addition scenarios using single crystal monolayer and bulk FGT structures (Figs. 5c-j). In both scenarios, we found that the DMI is anisotropic in plane yielding $d_{[100]} \neq d_{[110]}$ (Figs. 5c-i). Of note, similar behavior was also reported for out-of-plane magnetized bcc Au/Co/W structures[45]. For O-substitution case, we find that the single crystal monolayer DMI is nonmonotonic as a function of oxygen concentration, being weakly anticlockwise (resp. strongly clockwise) for low (resp. high) concentrations (Figs. 5d-f). As for the case of O-addition scenario, the DMI strength monotonically increases as a function of oxygen concentration, although $d_{[100]}$ and $d_{[110]}$ configurations give opposite DMI chirality (Figs. 5g-i). Regarding the DMI in the bulk O-substituted FGT structures, very importantly, we found additional DMI contributions arising from the proximity of pure FGT cell with the oxidized layer O-FGT. For instance, it follows from $\Delta E_{SOC}$ distribution shown in Fig. 5j, that O-FGT and FGT parts of this bulk structure provide clockwise [100] DMIs with -0.6 meV and -0.8 meV contributions, respectively, resulting in total value of -1.4 meV indicated by black solid square in Fig. 5f. The difference of about -1 meV between the total DMI values for the bulk FGT/O-FGT and O-FGT fully oxidized single crystal monolayer (open square in Fig. 5f), clearly indicates the large clockwise DMI contribution associated with the bulk FGT part in this structure. Similar conclusions can be deduced for [110] structure. As for O-addition scenario, these net FGT bulk contributions in FGT/O-FGT resulting from oxygen gradient within the structure are significantly smaller (Fig. 5i).

  Using these theoretical findings, we can now analyze the resulting DMI in our samples supposing that these oxidation scenarios occur within interfacial areas with transient Te-O concentration represented by shaded areas in Fig. 5b, with top (5.9 nm) oxidized region being thicker compared to the bottom one (3.3 nm). These relatively thick regions with variable oxidation rate within them suggest that the DMI is not "localized" at narrow interfaces between atomic layers.

Instead, the whole thickness regime with a finite oxidation gradient serves as DMI-enhancing layer across few nanometers, which works together with Heisenberg exchange, dipolar energy and anisotropy, leading to the formation of chiral magnetic skyrmions and their lattices observed here. Although two FGT/O-FGT interfaces are symmetrically present in our FGT sample and therefore may counter act, the magnitudes of DMI in these interfaces may be very different due to largely asymmetric oxidation profile and aforementioned scenarios (substitution and addition). In fact, even in case of only substitution scenario present, the overall net clockwise DMI will be present due to oxidation region asymmetry. Moreover, both O and Te interfacial gradients favors O-substitution scenario which gives rise to clockwise DMI provided by FGT adjacent layers.

To further corroborate our ab-initio study, we performed magnetic imaging of another FGT-based heterostructure without any O-FGT interfaces, as shown in Supplementary Fig. S11. In this experiment, we used DPCM by configuring the magnetic structure in low-magnification scanning mode (STEM), which enables the direct measurement of the in-plane magnetic components in focus to avoid image blurring which often occurs in Fresnel-LTEM imaging obtained at defocused mode[46]. The DPCM measurement and analysis present that the observed magnetic textures are Bloch-type, in good agreement with previous LTEM results observed in a FGT flake without oxidized interfaces[47]. This is significant, as it highlights that we can now manipulate the chiral states of magnetic texture (i.e. Néel-type vs. Bloch-type) in vdW heterostructures by employing different fabrication processes. Moreover, following the physical insights revealed by our work, we believe that future studies could utilize 2D heterostructures consist of non-oxidized FGT and other oxide vdW materials (e.g. FGT/$V_2O_5$) for engineering interfacial DMI in a more controllable way.

Nevertheless, we believe further systematic experimental studies probing the dependence of spin textures on the total FGT thickness, and/or the internal magnetization profile of skyrmions from top to bottom layers in FGT considering the role of van der Waals interactions could shed light into more precise tailoring of DMI and resulting magnetic textures in FGT crystal and heterostructures. After the initial submission of this manuscript we became aware of similar studies on skyrmion-like spin textures in the same (FGT) or similar (e.g. $Cr_2Ge_2Te_6$) two-dimensional van der Waals material by Wang *et al.*[43], Ding *et al.*[47], Han *et al.*[48] and Wu *et al.*[49]

In summary, we observed Neél-type chiral magnetic skyrmions and their lattice phase stabilization in a vdW ferromagnet FGT using high resolution magnetic microscopy. We examined the stability of SkX in FGT over a wide range of temperatures and magnetic fields, including its zero-field manifestation. We also demonstrated the current-driven motion of individual skyrmion in FGT-based vdW heterostructure, highlighting its potential for future skyrmionics devices. We performed symmetry analysis and first principles calculations supported by additional experiments to unveil the origins of the emergent Néel-type spin textures, namely DMI at the oxidized interfaces of FGT, which also demonstrates the controllability of chiral states in vdW heterostructures by process and/or interfacial material engineering. The possibility to achieve and electrically manipulate magnetic skyrmions in vdW magnets marks a significant advance in vdW magnet-based spintronics. Along with the large potential of skyrmions for future spintronic devices to store, process, and transmit data with extremely low power cost, this work will pave a new route towards vdW magnet-based topological magnetism and skyrmion-electronics.

**Acknowledgments.**
S.W. acknowledges the support from IBM Research and the managerial support from Guohan Hu and Daniel Worledge. S.W. also acknowledges Jiadong Zang for reading this manuscript and providing helpful comments. X.Z.Y acknowledges the support from Grants-In-Aid for Scientific Research (A) (Grant No. 19H00660) from Japan Society for the Promotion of Science (JSPS). T.-E.P., S.J.K., K.M.S., K.K., J.C. and H.C.K. acknowledge the support from the KIST Institutional Program (2E30600) and the National Research Council of Science and Technology (NST) (Grant no. CAP-16-01-KIST) by the Korean government (MSIP). K.K. acknowledges the support from the Basic Research Laboratory Program through the National Research Foundation of Korea (NRF) funded by the MSIT (NRF-2018R1A4A1020696). X.Z. was supported by the Guangdong Basic and Applied Basic Research Fund (Grant No. 2019A1515110713), and the Presidential Postdoctoral Fellowship of The Chinese University of Hong Kong, Shenzhen (CUHKSZ). Y.Z. acknowledges the support by the President's Fund of CUHKSZ, Longgang Key Laboratory of Applied Spintronics, National Natural Science Foundation of China (Grant No. 11974298), and Shenzhen Fundamental Research Fund (Grant No. JCYJ20170410171958839) and Shenzhen Peacock Group Plan (Grant No. KQTD20180413181702403). M.E. acknowledges the support by the Grants-in-Aid for Scientific Research from JSPS KAKENHI (Grant Nos. JP18H03676, JP17K05490 and JP15H05854) and also the support by CREST, JST (Grant Nos. JPMJCR16F1 and JPMJCR1874). X.L. acknowledges the support by the Grants-in-Aid for Scientific Research from JSPS KAKENHI (Grant Nos. 17K19074, 26600041 and 22360122). J.C acknowledges the support of Yonsei-KIST Convergence Research Institute. Y.D.K was supported by Samsung Research & Incubation Funding Center of Samsung Electronics under Project Number SRFC-TB1803-04 and grant from Kyung Hee University in 2018 (No. KHU-20181299). A. H. and M. C. acknowledge support from European Union's Horizon 2020 research and innovation programme under grant agreements No. 696656 and 785219 (Graphene Flagship). J.L and H.Y. acknowledge support from the National Natural Science Foundation of China (11874059) and Zhejiang Province Natural Science Foundation of China (LR19A040002). Part of this work was performed at the MAXYMUS endstation at Berlin Electron Storage Ring Society for Synchrotron Radiation II (BESSYII). We thank HZB for the allocation of neutron/synchrotron radiation beamtime. Part of this work was also performed at the PolLux (X07DA) endstation of the Swiss Light Source, Paul Scherrer Institut (PSI), Villigen, Switzerland. We thank PSI for the allocation of synchrotron radiation beamtime.

**Author contributions.**
S.W. designed and conceived the study. T.-E.P. prepared films, fabricated devices and performed device characterizations with the support from S.J.K. K.M.S., K.K. and Y.D.K.. T.-E.P, K.M.S., K.K., M.W., S.F., J.R. and S.W. performed STXM experiments at BESSY II in Berlin, Germany and at Swiss Light Source in Villigen, Switzerland. L.P. and X.Z.Y. performed Lorentz-TEM experiments and analyzed the data. F.S.Y. performed DPCM and Lorentz-TEM experiments on non-oxidized FGT and analyzed the data. J.L., A.H., A.F., M.C. and H.Y. performed the ab initio calculations on DMI in FGT crystal, and analyzed the results. X.Z., J.X., Y.Z., M.E. and X.L. provided symmetry analysis on DMI in FGT crystal. T.-E.P. drafted and L. P., X.Z., X.Z.Y., F.S.Y. and S.W. revised the manuscript and all authors reviewed the manuscript.

**Competing Interests**. The authors declare no competing interests.

**Figures**

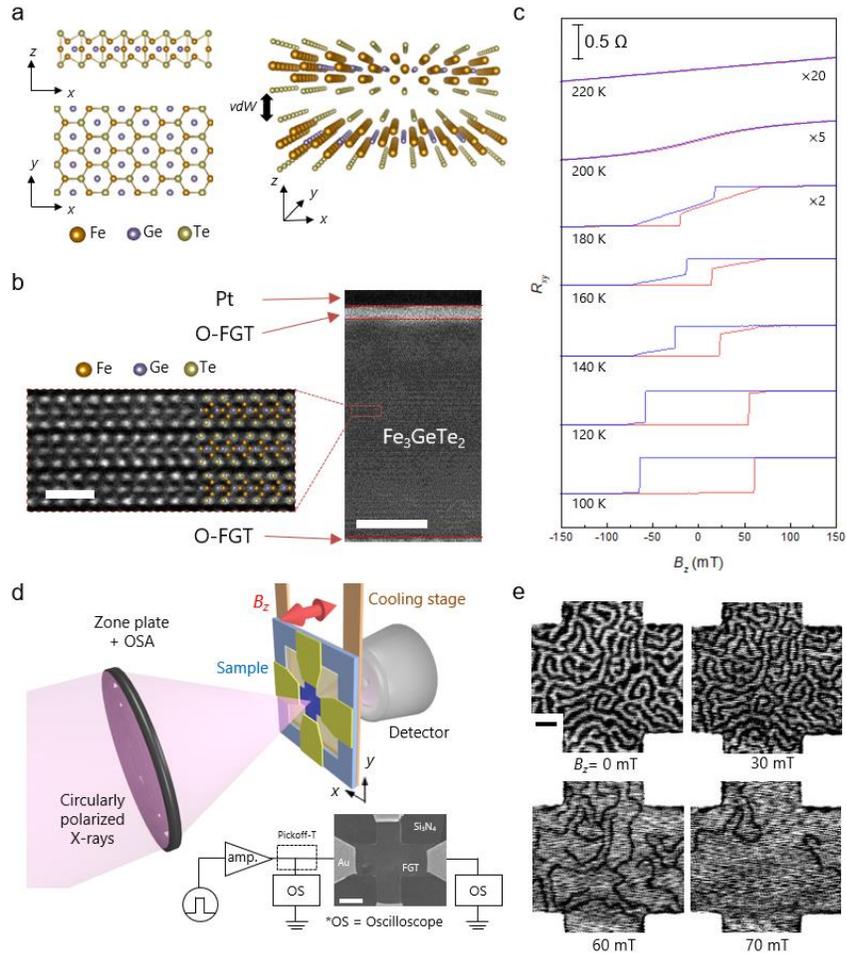

**Fig. 1 | Crystal structure, Hall measurement and the X-ray microscopy imaging of domain structures of a van der Waals Fe₃GeTe₂. a,** An atomic structure of a Fe$_3$GeTe$_2$ (FGT) monolayer (left) and the structure of a FGT bi-layer with an interlayer van der Waals (vdW) gap (right). Fe$^{III}$ and Fe$^{II}$ represent the two inequivalent Fe sites in the +3 and +2 valence states, respectively. **b,** Cross-sectional high-resolution transmission electron microscopy (HRTEM) image of the FGT with a Pt capping Hall-bar device fabricated on 100 nm-thick Si$_3$N$_4$ membrane substrate. (right, scale bar is 20 nm) Oxidized FGT is indicated as O-FGT. The enlarged panel shows the high angle annular dark field (HAADF) image in scanning TEM mode of the red-dashed highlighted area in **b**. (left, scale bar is 1 nm). **c,** Temperature dependent Hall resistance ($R_{xy}$) as a function of applied out-of-plane magnetic field, $B_z$. **d,** Schematic of scanning transmission X-ray microscopy (STXM) experimental setup used for magnetic domain imaging and simultaneous electrical pulse injections. The inset shows scanning electron microscopy (SEM) image of the measured device with Hall bar geometry. Scale bar, 4 μm. Two electrode pads on horizontal *x*-axis were used for electrical pulse applications, and oscilloscopes before and after device were used to verify the pulse profiles before and after device, respectively. **e,** Exemplary STXM images acquired as a function of increasing magnetic field from $B_z = 0$ mT to $B_z = 80$ mT at 120 K. Dark and bright contrast correspond to magnetization of Fe atoms oriented down (-$M_z$) and up (+$M_z$), respectively. Scale bar, 1 μm.

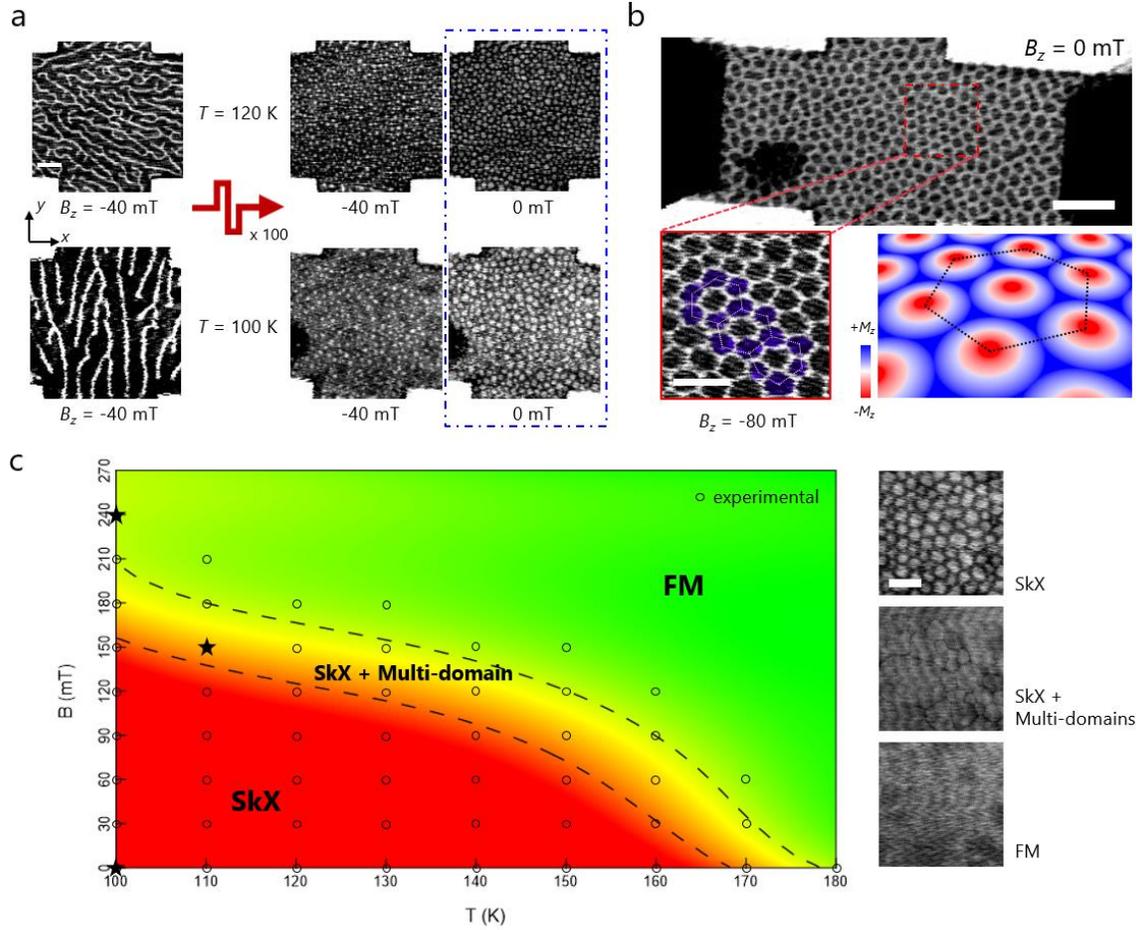

**Fig. 2 | Generation and stabilization of magnetic skyrmion lattice phase. a,** The two images on the left side were acquired at $B_z$ = -40 mT at 120 K and 100 K after the initial saturation at $B_z$ = +200 mT, respectively, where an initial labyrinth domain states were stabilized. The right two images at $B_z$ = -40 mT were acquired after the application of bipolar pulse bursts at 120 K and 100 K, respectively, and the other two images at $B_z$ = 0 mT were acquired after removing magnetic fields. Scale bar, 1 μm. **b,** Representative STXM image of skyrmion crystal (SkX) stabilized over the whole FGT device at $B_z$ = 0 mT and $T$ = 100 K. Scale bar, 2 μm. For clarity, the enlarged image of SkX was obtained at $B_z$ = -80 mT. Scale bar, 1 μm. The hexagonal white lines are drawn to guide eye for the ordered SkX, and the inset schematic represents the exemplary magnetic configuration of SkX found in chiral magnets for comparison. Note that skyrmion polarity in **b** (-$M_z$ core) is different from **a** (+$M_z$ core), as the initial field-sweep procedure of reversed field direction was used before the pulse application: $B_z$ = -200 mT → +40 mT. **c,** Experimental phase diagram of magnetic configurations as a function of temperature and magnetic field. Experimentally measured positions are marked with open circles, and star symbols correspond to exemplary images shown on the right side of the phase diagram. Three representative images show each magnetic configuration state: SkX, SkX + multi-domains, and saturated ferromagnet (FM). Scale bar, 1 μm. Black dashed lines in phase diagram are guide to the eyes to indicate the phase boundaries.

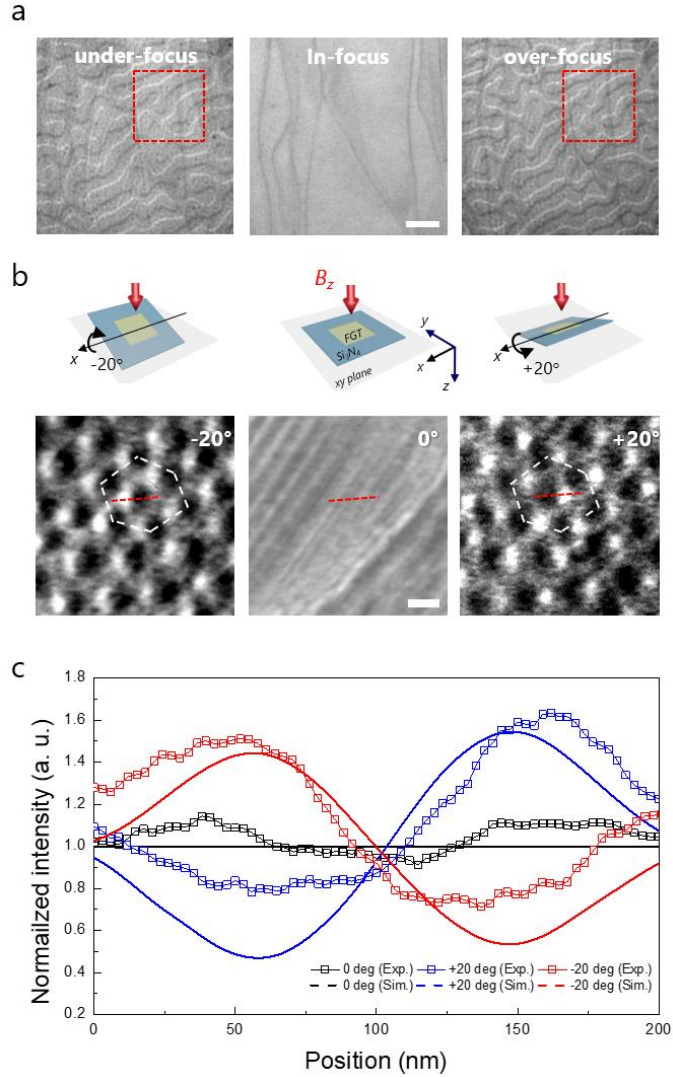

**Fig. 3 | Lorentz transmission electron microscopy (LTEM) measurements of skyrmion crystals (SkX). a,** Lorentz transmission electron microscopy (LTEM) images of magnetic configurations at under-focus (left), in-focus (middle) and over-focus (right) acquired at zero field and 160 K, respectively. Scale bar, 500 nm. **b,** LTEM images of magnetic skyrmion crystal (SkX) taken at the sample tilting angle of -20° (left), 0° (middle) and 20° (right) with respect to $x$-axis at $B_z$ = -40 mT and 160 K, respectively. The defocus values are ±3 mm for observing over/under focus image. The strip-like contrasts in **b** at zero-tilt angle indicate bending contours of the flake. White hexagonal-shaped lines are eye guides for the unit of the SkX. Scale bar, 200 nm. Upper panels in **b** are schematic drawings of the flake orientations with respect to the incident electron beam. Solid red lines in **b** guide the line-scan positions for contrast profiles. **c,** Experimentally measured (symbols) and simulated (solid lines) contrast profiles across a single skyrmion as shown in **b**. Note that the simulated profiles were obtained for a 100 nm-size Néel-type skyrmion.

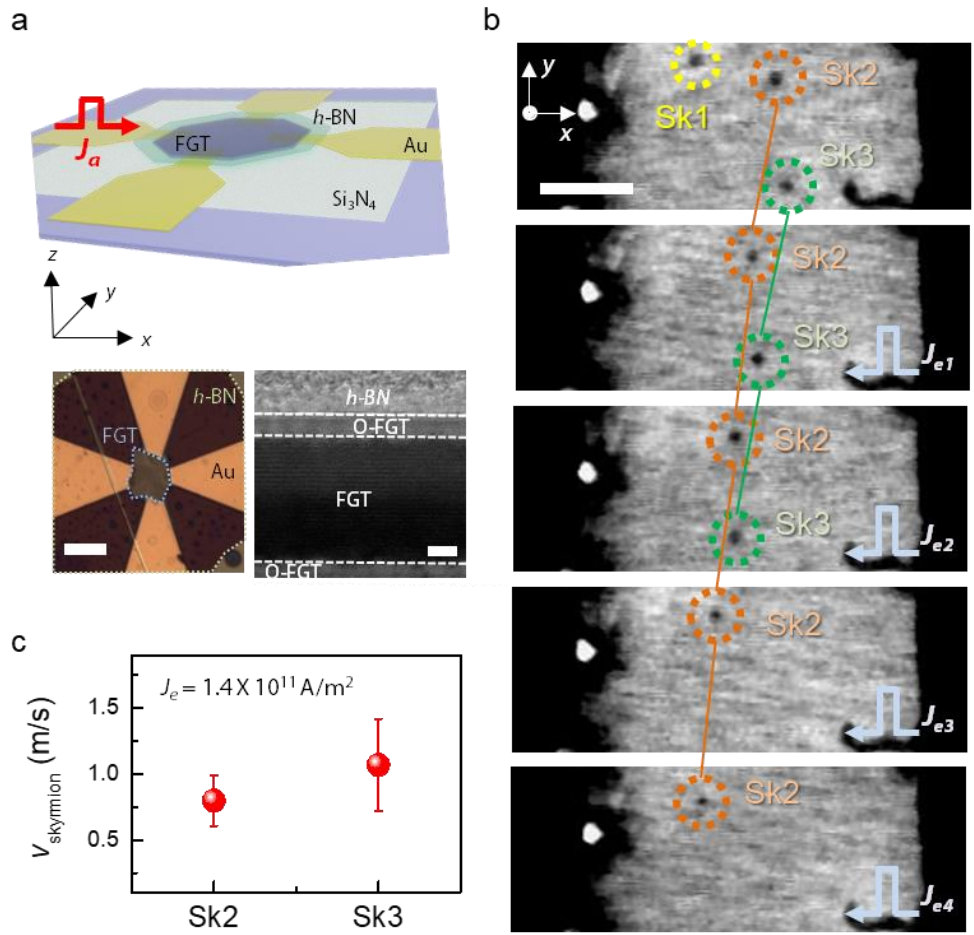

**Fig. 4 | Current-driven motion of skyrmions. a,** Schematic image of the FGT device used for the experiment. Two bottom inset images show the optical microscopy image of a FGT device capped by *h*-BN used for the STXM experiment (left, scale bar is 10 μm), and the cross-sectional high-resolution transmission electron microscopy (HRTEM) image of the device (right, scale bar is 10 nm). **b,** Sequential STXM images showing skyrmion configurations in the FGT device, where each STXM image was acquired after injecting 5 unipolar current pulses along the +*x* direction (electron flow to -*x* direction), with an amplitude $J_e = 1.4\times10^{11}$ A/m$^2$ and duration 50 ns. All images were obtained at the oblique field of -20 mT and 100 K. Individual skyrmions are outlined in colored circles for clarity (Sk1, Sk2, and Sk3). Scale bar, 1 μm. **c,** The average velocities for two representative skyrmions, Sk2 and Sk3, where error bars denote the standard deviation of multiple measurements.

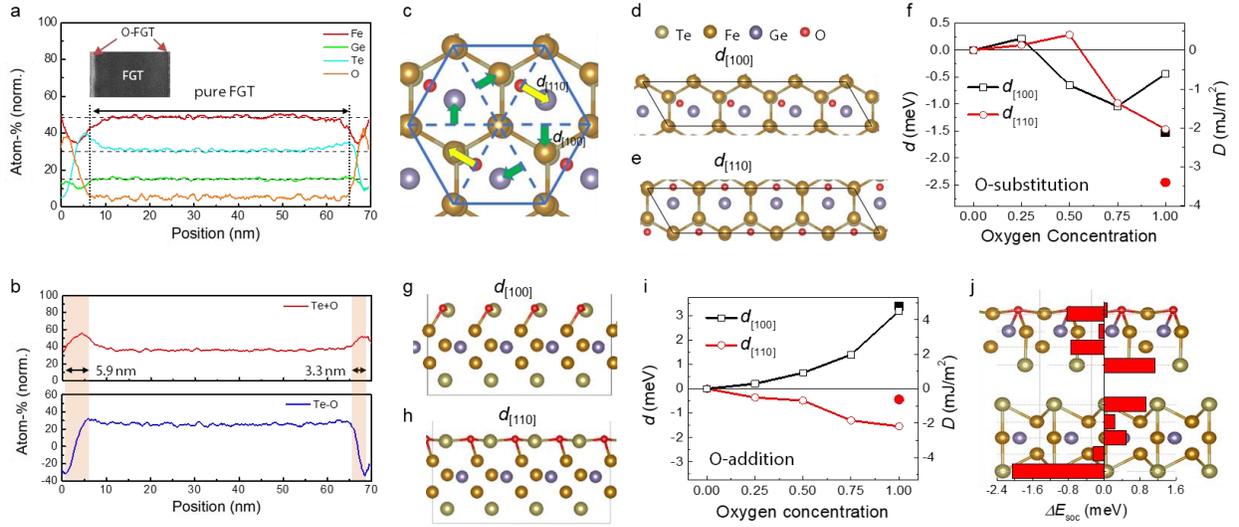

**Fig. 5 | First principle calculation of Dzyaloshinskii-Moriya interaction (DMI) in FGT crystal and interfaces. a,** Atomic concentration distribution of Fe, Ge, Te and O atoms across sample thickness within the FGT crystal used for STXM measurements. **b**, Relative distribution of Te and O atomic concentrations, i.e. their sum and difference across the FGT sample thickness. **c,** Top views of relaxed crystal of oxygen-substituted $Fe_3GeTe_{2-x}O_x$ (x = 1) relaxed structure showing DMI vectors in **d** [100] and **e** [110] directions. **f**, The calculated DMI parameters in [100] and [110] in-plane directions as a function of oxygen substitution concentration of $Fe_3GeTe_{2-x}O_x$ monolayer (x = 0, 0.25, 0.5, 0.75, 1). Solid symbols show the DMI values for FGT/O-FGT bulk structures. **g** and **h**, The side view of oxygen-added FGT structures along [100] and [110] directions. **i,** The same as **f** for oxygen addition concentration. **j,** Side view of bulk FGT [100]/O-FGT structure and layer-resolved SOC energy difference, $\varDelta E_{SOC}$, associated with DMI distribution.